\documentclass[aps,nofootinbib]{revtex4}

\usepackage{graphicx}% Include figure files
\usepackage{dcolumn}% Align table columns on decimal point
\usepackage{bm}% bold math
\usepackage{amssymb,amsfonts}
\usepackage{epsfig}
\usepackage{epstopdf}
\usepackage[all]{xy}
\usepackage{amsthm}
\usepackage{dcolumn}
\usepackage{hyperref}
\usepackage{url}
\usepackage{textcomp}
\usepackage{csquotes}
\usepackage{dsfont}

\newcommand{\be}{\begin{equation}}
\newcommand{\ee}{\end{equation}}
\newcommand{\bea}{\begin{eqnarray}}

\newcommand{\eea}{\end{eqnarray}}

\def\inbar{\,\vrule height1.5ex width.4pt depth0pt}
\def\IR{\relax{\rm I\kern-.18em R}}
\def\IC{\relax\hbox{$\inbar\kern-.3em{\rm C}$}}

\begin{document}

\title{On the graviton two-point function and its infrared behavior in de Sitter spacetime}

\author{Surena Rahbardehghan\footnote{sur.rahbardehghan.yrec@iauctb.ac.ir}} \author{Hamed Pejhan\footnote{h.pejhan@piau.ac.ir}}
\affiliation{Department of Physics, Science and Research Branch, Islamic Azad University, Tehran, Iran}

\begin{abstract}
In our previous paper (Phys. Rev. D 94, 104030 (2016)), inspired by the work of Allen and Turyn (Nucl. Phys. B 292, 813 (1987)), we expressed the graviton two-point function in terms of maximally symmetric bitensors in de Sitter spacetime. Quite contrary to the result of their work, however, we explicitly showed that the de Sitter symmetry breaking is universal and the associated infrared divergences cannot be gauged away. In this Letter, the origin of this contradiction, which is greatly related to the well-known scientific dispute about the infrared behavior of the graviton two-point function, is clarified.
\end{abstract}

\maketitle
%\tableofcontents
%\include{intro}

Due to the relevance of de Sitter (dS) spacetime to the inflationary epoch of the early Universe and also its current accelerated expansion, investigating into the graviton two-point function and its infrared (IR) behavior on this background are of particular significance and have been performed extensively in the literature from various point of views. Considering the de Sitter global coordinate, the Euclidean approach of Allen and Turyn \cite{allen2} is one of the most striking approaches to the dS graviton two-point function that provides a fundamental context upon which many similar works have been accomplished till today (see for instance \cite{higuchi1,higuchi2} and many other papers in the period of time between them). Through this approach, assuming that the Euclidean vacuum being de Sitter invariant, the graviton two-point function is written in terms of maximally symmetric bitensors.\footnote{Bitensors are functions of two points $(x, x')$ and behave like tensors under coordinate transformations at each point. They are called maximally symmetric if they respect dS invariance \cite{allen1}.} On this basis, it has been argued that the infrared behavior of the graviton two-point function is highly gauge-dependent and can be eliminated by a proper gauge-fixing procedure. Here, it must be underlined that this method is one of the most fundamental reasonings in the stream that strongly asserts the gauge dependency of the infrared behavior of the dS graviton field in the well-known controversy (see \cite{Pejhan2} and references therein) about the dS graviton field IR behavior.\\

In this Letter, we insist on the crucial fact that only on the foundation that the Euclidean vacuum being de Sitter invariant, the two-point functions can be expressed in terms of maximally symmetric bitensors. Otherwise, if the Euclidean vacuum does not satisfy the dS invariance condition, any attempt to write the two-point function in terms of maximally symmetric bitensors and analyze its IR behavior is pointless. Here, we particulary emphasize the de Sitter invariance aspect which should be understood in the sense of the action of the de Sitter group. In this regard, in \cite{Pejhan2,Pejhan1}, considering the ambient space notation\footnote{Describing the dS spacetime as a one-sheeted hyperboloid embedded in a five-dimensional Minkowski spacetime which makes apparent the group theoretical content of the considered model, contrary to a more compact intrinsic formalism \cite{Banerjee}.}, the solution to the linearized Einstein equation in de Sitter global coordinate is examined through a rigorous group theoretical approach. The main result of this study is that, under the action of the de Sitter group, the solutions do not establish a closed set for any gauge field. Actually, in order to preserve the dS invariance, the entire set of the negative-frequency solutions to the graviton field (regarding the conformal time) must be included \cite{Pejhan2,Pejhan1}. More exactly, the Euclidean vacuum is not dS-invariant.\\

Obviously, due to the violation of the dS invariance, the Euclidean approach cannot be applied to the dS graviton field from the very beginning and therefore no maximally symmetric dS graviton two-point function exists at all. As a result, any argument in this context about infrared behavior of the graviton two-point function (see for instance \cite{allen2,higuchi1,higuchi2}) is unreliable. Here, we must note that, in our previous work \cite{Pejhan2}, respecting the dS invariance requirement for evaluating the graviton two-point function in terms of maximally symmetric bitensors, we first constructed the smallest complete, nondegenerate, and invariant inner product space for the graviton field by including all the negative-frequency solutions. Then, possessing the invariant vacuum state (the Krein vacuum state), inspired by the work of Allen and Turyn, we obtained the associated maximally symmetric Krein graviton two-point function. In fact, the only maximally symmetric two-point function normally arose was the Krein two-point function which is actually the commutator; it is not of positive type, it does not permit to choose the physical states, and also it unavoidably breaks the analyticity of the theory \cite{Pejhan2,Pejhan1}. Moreover, it is interestingly free of any IR divergences \cite{Pejhan2,Pejhan1}.\\

We hope that this study would clarify the contradiction between our results and the Euclidean approach's and shed the light on the ambiguity about the IR behavior associated with the graviton field in de Sitter spacetime.\\

\section*{Acknowledgements}
We would like to thank Professor R.P. Woodard for his instrumental comments.

\end{document}